\documentclass{aastex631}
\usepackage[caption=false]{subfig}
\usepackage{amsmath}
\turnoffedit

\accepted{29-Dec-2021}
\submitjournal{ApJ}

\begin{document}
\title{ A Tight Three-parameter Correlation and Related Classification on Gamma-Ray Bursts }

\author[0000-0003-2413-9587]{Shuai Zhang }
\affiliation{Department of Space Sciences and Astronomy, Hebei Normal University, Shijiazhuang 050024, China.}

\author[0000-0001-8876-2357]{Lang Shao}
\affiliation{Department of Space Sciences and Astronomy, Hebei Normal University, Shijiazhuang 050024, China.}
 
\author[0000-0003-4111-5958]{Bin-Bin Zhang}
\affiliation{School of Astronomy and Space Science, Nanjing University, 210093 Nanjing, China.}

\author{Jin-Hang Zou}
\affiliation{Department of Space Sciences and Astronomy, Hebei Normal University, Shijiazhuang 050024, China.}
\affiliation{School of Astronomy and Space Science, Nanjing University, 210093 Nanjing, China.}

\author{Hai-Yuan Sun}
\affiliation{Department of Space Sciences and Astronomy, Hebei Normal University, Shijiazhuang 050024, China.}

\author{Yu-Jie Yao}
\affiliation{Department of Space Sciences and Astronomy, Hebei Normal University, Shijiazhuang 050024, China.}

\author[0000-0003-1454-2268]{Lin-Lin Li}
\affiliation{Department of Space Sciences and Astronomy, Hebei Normal University, Shijiazhuang 050024, China.}

\email{zhangshuai@hebtu.edu.cn; lshao@hebtu.edu.cn}

\begin{abstract}
Gamma-ray bursts (GRBs) are widely believed to be from massive collapsars and/or compact binary mergers, which accordingly, would generate long and short GRBs, respectively. The details on this classification scheme have been in constant debate given more and more observational data available to us. In this work, we apply a series of data mining methods to studying the potential classification information contained in the prompt emission of GRBs detected by the Fermi Gamma-ray Burst Monitor. A tight global correlation is found between fluence ($f$), peak flux ($F$) and prompt duration ($T_{90}$) which takes the form of {$ \log {\it f}= 0.75 \log T_{90} +0.92 \log F -7.14$}. Based on this correlation, we can define a new parameter $L = 1.66\log T_{90} + 0.84 \log {\it f} - 0.46 \log F + 3.24$ by linear discriminant analysis that would distinguish between long and short GRBs with much less ambiguity than $T_{90}$. 
We also discussed the three subclasses scheme of GRB classification derived from clusters analysis based on a Gaussian mixture model, and suggest that, besides SGRBs, LGRBs may be divided into long-bright gamma-ray bursts (LBGRBs) and long-faint gamma-ray bursts (LFGRBs), LBGRBs have statistical higher $f$ and $F$ than LFGRBs; further statistical analysis found that LBGRBs also have higher number of GRB pulses than LFGRBs.

\end{abstract}
\keywords{Gamma-ray bursts (629); Multivariate analysis (1913); Classification (1907);}

\section{Introduction}
Gamma-ray bursts (GRBs) are the most violent explosions from the deep universe and have been studied extensively ever since their discoveries in the late 1960s \citep{1973ApJ...182L..85K}. 
They are generally considered to be composed of short GRBs, with durations shorter than 2s, and long GRBs, which last longer \citep{1993ApJ...413L.101K}. The long GRBs are most likely powered by the collapse of a (rapidly rotating) massive star often associated with a bright supernova \citep{1999ApJ...524..262M,2006ARA&A..44..507W}, while the short GRBs are expected to be from coalescence of compact object binaries involving at least one neutron star \citep{1989Natur.340..126E,1991AcA....41..257P,1992ApJ...395L..83N}. 

There seem to be obvious differences in emission characteristics between long and short GRBs, for example, short GRBs have statistically larger hardness ratios than the long GRBs \citep[e.g.,][]{2000AIPC..526..277K}. In addition, the host galaxies of long GRBs also have different features from those of short GRBs. Long GRBs whose progenitors are young massive stars occur in star-forming galaxies \citep{2002AJ....123.1111B,2006Natur.441..463F,2011MNRAS.412.1473L} while the host galaxies of short GRBs include both early-type and star-forming galaxies \citep{2005Natur.438..988B,2005Natur.437..845F,2006ApJ...638..354B,2011ApJ...730...26F,2013ApJ...776...18F}. The host galaxies of long GRBs also have more concentrated and smaller age distributions, smaller stellar mass \citep{2009ApJ...691..182S,2010ApJ...725.1202L}, and lower metal abundance \citep{2004ApJ...617..240K,2004ApJ...613..898T}. The two subclasses of GRBs also have different offset (distance from the center of the host galaxy) distribution \citep{2013ApJ...776...18F}. \cite{2017ApJ...844...55Z} have found a correlation between the luminosity of short GRB afterglows and the offset from host galaxies, while for long GRBs, neither prompt emission nor afterglow satisfied a similar correlation with its offset.

In spite of the many differences proposed between long and short GRBs, there is not a clear-cut observational standard to strictly distinguish the two types of bursts. In addition, it is not very certain whether GRBs can only be divided into two subclasses or there should be more subclasses. Over the years, researchers have proposed different classification scenarios, such as introducing a medium duration subclasse between long and short GRBs based on $T_{90}$ \citep{1998ApJ...508..757H,1998ApJ...508..314M,2000ApJ...538..165H,2010ApJ...713..552H}, or classifying GRBs by defining and applying variable parameters \citep{2010ApJ...725.1965L}. \citet{2016ApJS..227....7L} conducted a comprehensive study of the comparison between long and short GRBs by analyzing 14 relevant parameters in detail and came up with three best parameters for the classification purpose ($T_{90}$,$f_{\rm eff}$, $F_{\rm light}$). However, no single parameter alone is yet good enough to accurately determine the category for a particular burst. Therefore, multiple-parameter criteria for physical classification seem to be unavoidable.

The development of machine learning and data mining technology provide us with new research tools. Traditional machine learning methods such as principle component analysis (PCA) and ${\it k}-$th nearest neighbor, etc., can be used to perform spectral line certification in the spectrum and eliminate background radiation \citep{2017A&A...600A..54M,2016arXiv161207549S}. \citet{2016MNRAS.458.3821U} estimated the redshift of GRBs by using random forest algorithm.

In terms of GRBs classification, there have also been some related applications.  With new methods adopted, however, some new and different results or conclusions about GRB classification have been proposed. \citet{2017arXiv170805668B} analyzed two-dimensional clustering of GRBs using duration and hardness and obtained different classification conclusions of either two or three components for different detectors or datasets. \citet{2019JPhCS1400b2010S} analyzed GRBs detected by Konus-Wind with similar methods and also suggested a three-component classification. \citet{2017MNRAS.469.3374C,2018MNRAS.481.3196C} divided GRBs into five categories by gaussian (and t-)mixture-model-based cluster analysis, and \citet{2019Ap&SS.364..105H} proposed a three-subclass GRBs classification scheme. Meanwhile, some researches \citep{2019ApJ...870..105T,2019ApJ...887...97T} imply that the third class is unlikely to be a real phenomenon. Briefly, either the number of categories is different, or the characteristics of the subclasses are different. Thus the issues of GRB classification are far from being solved.

\section{data and algorithm}\label{data}
In this work, we analyze the sample of 2989 GRBs detected by Fermi GBM detectors
The parameters included in our analysis are duration $T_{90}$ (during which $90\%$ of the burst fluence was accumulated, in units of seconds), fluence $f$ (flux integrated over the full burst duration in 10 -1000 keV energy band, in units of erg~cm$^{-2}$), and peak flux $F$ (the peak flux in 10 -1000 keV energy band on the timescale derived by Bayesian Block method \citep{2013ApJ...764..167S}, in units of photon~cm$^{-2}$~s$^{-1}$), which are publicly available from the Fermi GBM burst catalog \citep{2020ApJ...893...46V}. We consider the data as a $ 2989 \times 3$ matrix $\it \boldsymbol  X$, which has three columns consisting of $\log T_{90}$, $\log f$ and $\log F$, respectively. We will analyze the three-dimensional distribution of this data set. 

All the algorithms used in this work are mature and complete tools in many fields. We will not go into details about mathematical theory, and just briefly introduce the characteristics of these algorithms related to the problems studied in this work.

\subsection{Principle Component Analysis}\label{sec:PCA}

PCA is a method that can be used in exploratory data analysis \citep{Pearson1901, Hotelling1933a, Hotelling1933b}. It can be understood as a method of projecting data set to a new orthogonal coordinate system in which the axes are linear combinations of initial parameters. The new coordinate system has the same dimensions as initial data, the first axis indicates the first principal component, which is defined as a direction that maximizes the variance of the projected data, and the second axis indicates the second principle component, defined as a direction that maximizes the variance of the projected data in a subspace perpendicular to the first principal component and, by analogy, each principal component can be taken as a direction orthogonal to the its former principal components that maximizes the variance of the projected data. Briefly, the order of the axes is arranged in accordance with the contribution to the total variance. PCA is commonly used for dimensionality reduction by ignoring the components with small variances and retaining those with large variances, thus the major information of the data can be retained within as few dimensions as possible. 

\subsection{Linear Discriminant Analysis}\label{sec:LDA}

Linear discriminant analysis (LDA) is a method used in statistics or other fields to find linear combinations of features that characterize or separate two or more classes of objects or events \citep{Fisher1936}. LDA is very similar to PCA in terms of ideas. However, LDA attempts to distinguish different classes of data while PCA just reduces the dimensionality while retaining most of the information in the data and does not consider the differences between classes at all.

\subsection{Gaussian Mixture Model}\label{sec:GMM}

A mixture model is a probabilistic model for representing the presence of subclusters within an overall data set and without requiring any single parameter to independently determine sample attribution. A Gaussian mixture model (GMM) is a special mixture model that assumes a mixture of a finite number of Gaussian distributions with unknown parameters to explain the data distribution \citep{Dinov2008,2012ITIP...21.2481G}. Gaussian distributions have been adopted by many GRB classification studies \citep{2017MNRAS.469.3374C,2019JPhCS1400b2010S,2017arXiv170805668B,2019ApJ...870..105T,2019ApJ...887...97T}. In this work, we assume the distributions of each subclass of GRBs to be Gaussian in logarithmic space and use GMM to derive the properties of the overall population from those subclasses.

\begin{figure*}
	\begin{center}
		\includegraphics[width=400 pt]{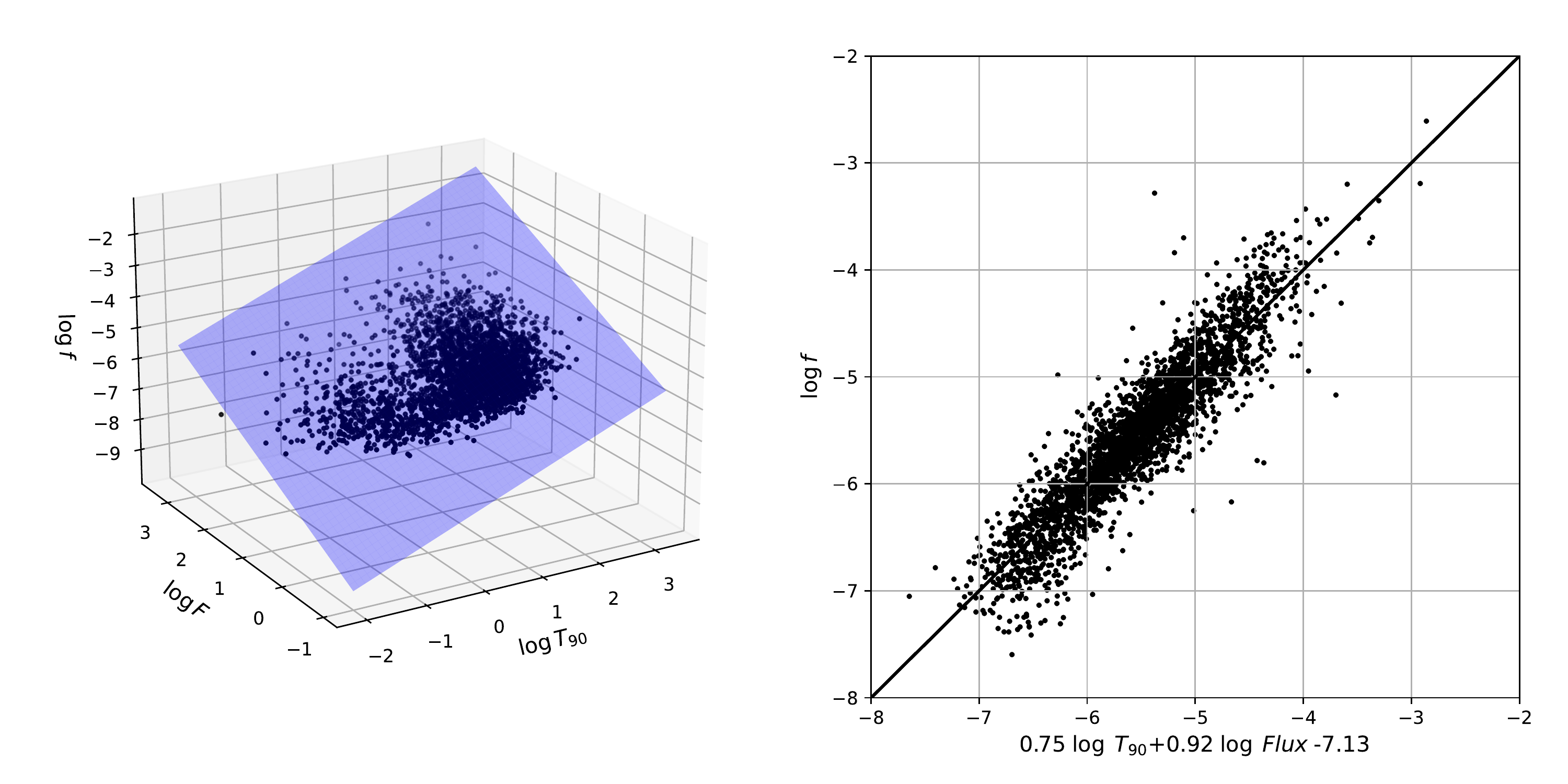}
		\caption{The left panel shows the correlation between $\log f$, $\log F$ and $\log T_{90}$ which is fitted by a plane in three-dimensional space: the blue plane is the fitted result. The right panel also shows the correlation. This 2D graph is a projection of the 3D graph on the left in a certain direction and this form is widely used in statistical research.}
		\label{fig1:corfit3p}
	\end{center}
\end{figure*}

\begin{figure}
	\begin{center}
		\includegraphics[width=200 pt]{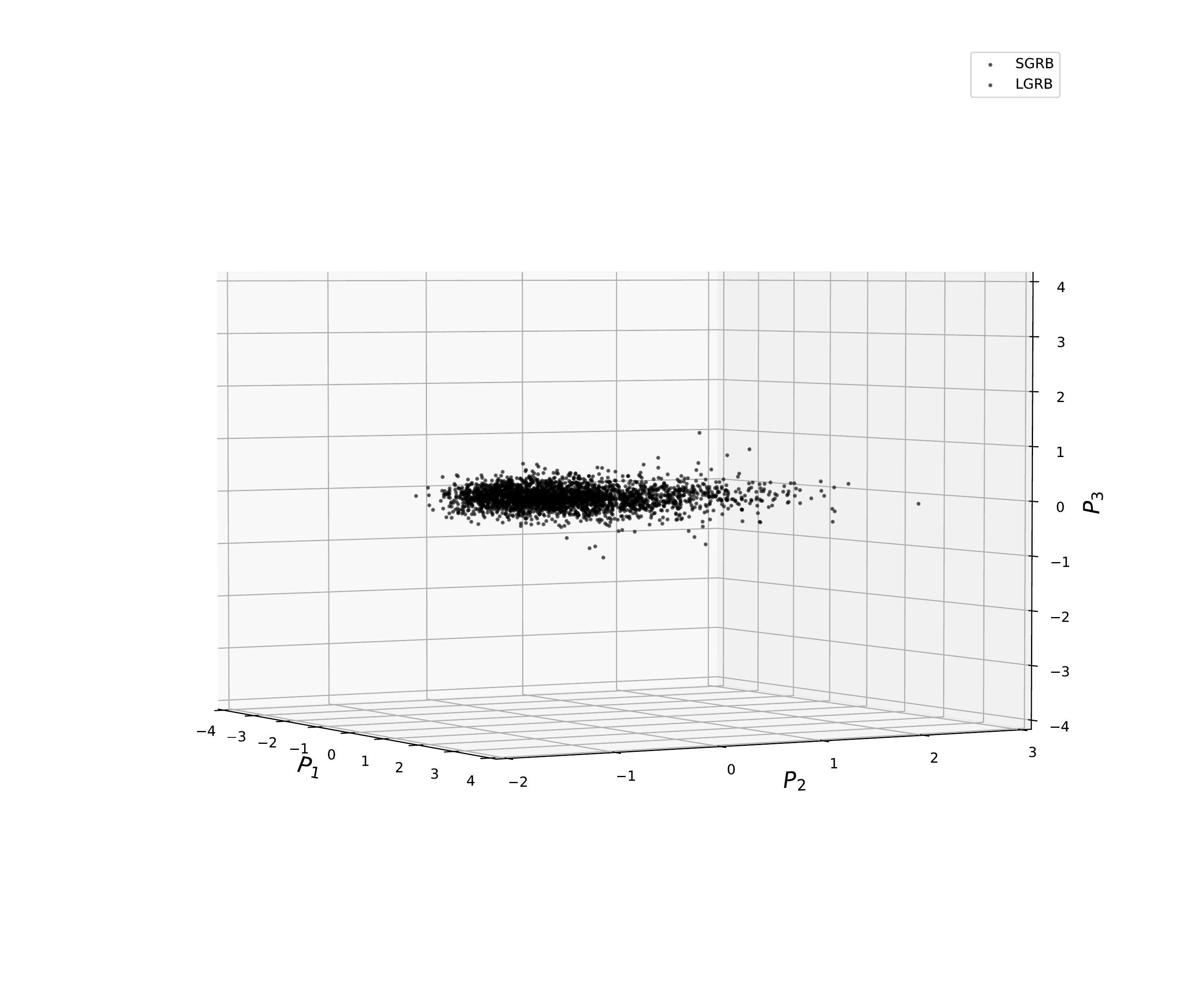}
		\caption{Distribution of GRBs in three-dimensional space. After principal component analysis, data was redrawn into the new space. The axes ($P_{1}$,$P_{2}$,$P_{3}$) are the three principle components concluded by PCA, which are linear combinations of $\log f$, $\log F$, and $\log T_{90}$. Obviously, the distribution of GRBs in the direction of $P_{3}$ is very concentrated, so we can retain most of the differences of information in the two-dimensional $P_{1}-P_{2}$ space.}
		\label{fig2:PCAtransf}
	\end{center}
\end{figure}

\section{Analysis}

\subsection{A Tight Three-parameter Correlation} 

As shown in the left panel of Fig.~\ref{fig1:corfit3p}, the three-dimensional distribution of the total fluence ($f$), peak flux ($F$) and duration $T_{90}$ concentrates very tightly in a certain plane of logarithmic scales. The best-fit function for this plane could be easily determined with a multivariate regression analysis as in:
\begin{equation}\label{eq:3pfit}
	\log {\it f}= 0.75 \log T_{90} +0.92 \log F -7.14
\end{equation}
It is a very tight global correlation for all GRBs in the sample with a correlation coefficient of $R=0.92$. A corresponding two-dimensional projection as shown in the right panel of Fig.~\ref{fig1:corfit3p} appears be more intuitive.

PCA is applied to the sample to compute the principal components and then transform the data into a new coordinate system that takes the principle components as the new axes. Accordingly, we get the principle components as in:
 
\begin{equation}
	{\it \boldsymbol P}=
	\begin{pmatrix}
		~0.75 & -0.44 & -0.48~\\
		~0.65 &  0.47 &  0.59~\\
		~0.04 &  0.76 & -0.64~\\
	\end{pmatrix}
\end{equation}

The initial sample variables $\it \boldsymbol X$ can be transformed into new coordinate system $\it \boldsymbol X^\prime$ by:
\begin{eqnarray}\label{eq:PCATransfor}
	{\it \boldsymbol X^\prime = (\it \boldsymbol X- \it \boldsymbol C) \cdot  \it \boldsymbol P }
\end{eqnarray}
where ${\it \boldsymbol C}= (1.12, -5.55, 0.81)$ is the center (judged by mean values) of three-dimensional distribution of initial sample variables $\it \boldsymbol X$. Therefore, we can express the new three principle components as:
\begin{eqnarray}
	{\it \boldsymbol X^\prime}&=&
	\begin{pmatrix}\nonumber
		P_1 \ P_2 \ P_3
	\end{pmatrix}\\
	&=&\begin{pmatrix}
		0.75\log T_{90} + 0.65 \log  {\it f} + 0.04 \log F +2.76 \\
		-0.44\log T_{90} + 0.47 \log  {\it f} + 0.76 \log F + 2.47 \\ 
		-0.48\log T_{90} + 0.59 \log  {\it f} - 0.64 \log F + 4.36
	\end{pmatrix}^{\rm T}
\end{eqnarray}

Three columns of the matrix $\it \boldsymbol X^\prime$ correspond to three principle components $P_1$, $P_2$, and $P_3$, respectively, which have been reverse sorted according to the variance. We can also evaluate the variance ratios ($\it \boldsymbol {Vr}$) of the three components as in:
\begin{eqnarray}
	{\it \boldsymbol {Vr}}  = \begin{pmatrix}
		Vr1 \quad Vr2 \quad Vr3
	\end{pmatrix} = \begin{pmatrix}
	0.70 \quad 0.27 \quad 0.02
\end{pmatrix}
\end{eqnarray}

The variance ratio of the third principle component is only $0.02$, which is almost negligible in comparison with that of the other two principle components. So the three-dimensional data can be reduced to two dimensions by dropping the third component. The updated distribution of our sample data in the new coordinate system $\it \boldsymbol X^\prime$ is shown in  Fig.~\ref{fig2:PCAtransf}.

From another perspective, the dropped component is not useless. Instead, it carries important consistency information of the data sample. In other words, the third principle component indicates a correlation between data parameters of $\log T_{90}, \log f $ and $\log F$. Eq.~\ref{eq:PCATransfor} indicates that the data sample should be centered at $\bf 0$ in the new coordinate system. So, from the third principle component, the mean value of data is:
\begin{equation}\label{eq:pca3p}
	-0.48\log T_{90} + 0.59 \log  {\it f} - 0.64 \log F + 4.36=0
\end{equation}
This is essentially a constraint on the data sample, which is a plane in the three-parameter space . Through equivalent transformation of the formula, we can see that Eq.~\ref{eq:pca3p} and Eq.~\ref{eq:3pfit} are almost the same.

To sum up, even though the dropped component in PCA rarely contains the difference information, it does contain the most consistent information of the sample. Later in this work, we will carry out analyses on the classification of GRBs based on the remaining components, i.e., the first and second principle components, which contain most of the difference information.

\subsection{Improved Classification Criteria}

\begin{figure*}
	\begin{center}
		\includegraphics[width=400 pt]{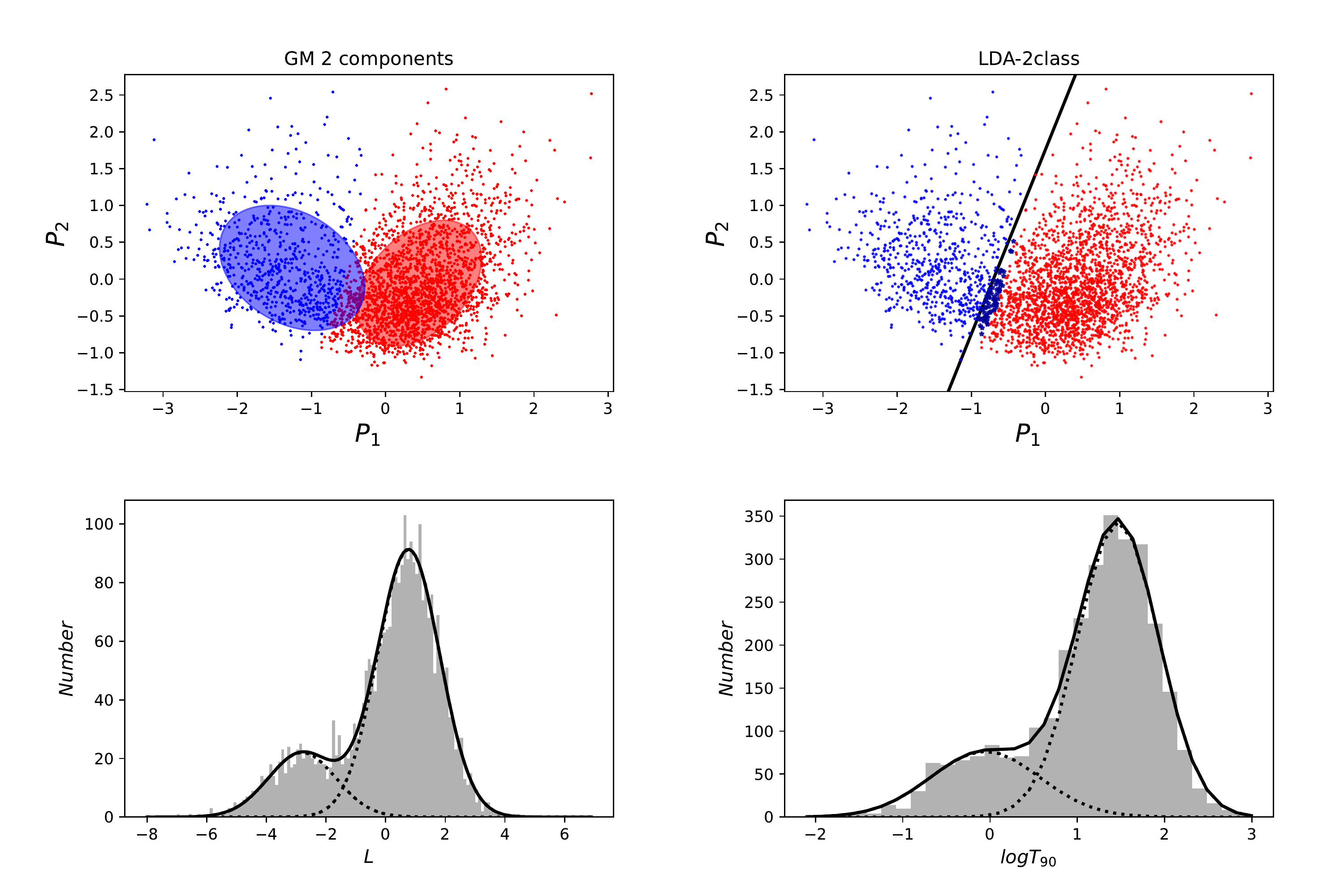}
		\caption{{\rm Upper left:} classification by two-component GMM. Blue dots represent short GRBs and red represent long GRBs, not strictly divided according to $ T_{90} $. The blue and red ellipses show the covariance of two Gaussian cluster components, respectively.
			{\rm Upper right:} reclassification through LDA based on GMM. The LDA classification result is only slightly different from that of GMM. GRBs whose category has changed are indicated by stars while the color still adopts the GMM scheme. 
			{\rm Lower left:} the projection distribution of the entire sample in the $L$ direction and also fitted by a two-component Gaussian function. The resolution of two Gaussian components is $R_{L} \approx 0.58$. 
			{\rm Lower right:} distribution of $\log T_{90} $; the entire sample(indicated by gray bars) is fitted by a two-component Gaussian function. The resolution of two Gaussian components is $R_{T} \approx 0.49$.}
		\label{fig3:LDA2class}
	\end{center}
\end{figure*}

Based on the result of PCA, we consider first and second principle components as a new data matrix :
\begin{eqnarray}\label{eq:newdata}
	{\it \boldsymbol {X_2}}&=&
	\begin{pmatrix}\nonumber
		P_1 \ P_2 
	\end{pmatrix}\\
	&=&\begin{pmatrix}
		0.75\log T_{90} + 0.65 \log  {\it f} + 0.04 \log F +2.76 \\
		-0.44\log  T_{90} + 0.47 \log  {\it f} + 0.76 \log F + 2.47 \\ 
	\end{pmatrix}^{\rm T}
\end{eqnarray}
This is obtained by removing the third component $P_3$ from $\it \boldsymbol X^\prime$. We use GMM to divide it into two subclasses. The result is shown in the upper left panel of Fig. \ref{fig3:LDA2class} where the two subclasses are represented in red and blue colors, respectively. In fact, the red and blue clusters are roughly composed of long and short GRBs, respectively, though the clusters are not single-handed chosen based on $T_{90}$. For convenience, we still use LGRB and SGRB to represent the red and blue data clusters, respectively. 

We then use the GMM classification result as the known training set, and use the LDA algorithm to find the optimal line that can distinguish the two subclasses, which is shown by an oblique line in the upper right panel of Fig.\ref{fig3:LDA2class}. For a small number of GRBs near the line, LDA gives a different class prediction from GMM. We can also reduce the result to one-dimensional space by transforming the diagonal line to a vertical one and projecting all data into the horizontal axis, as shown in lower left panel of Fig.\ref{fig3:LDA2class}. The conversion formula is:

\begin{eqnarray}
	L &=& \begin{pmatrix}
			P_1 \ P_2
		\end{pmatrix} \cdot  
		\begin{pmatrix}
			~1.80~\\
			~-0.70~\\
		\end{pmatrix} \\ \label{eq:LDATransfor}
	&=& 1.66\log T_{90} + 0.84 \log  {\it f} - 0.46 \log  F + 3.24, 
\end{eqnarray}
Where the constant $1.66$ just means a horizontal shift for the data points and has no meaning for the sample distribution characteristics. As for the classification, the important thing is the relative value of $L$, not the absolute value.

We use the new parameter $L$ defined by Eq. (\ref{eq:LDATransfor}), which is combination of three observational parameters, as the improved criteria to distinguish short and long GRBs. As shown in the lower left panel of Fig.\ref{fig3:LDA2class}, there is a significant bimodal structure. For comparison, we also fit $T_{90}$ in the lower right panel of Fig.~\ref{fig3:LDA2class} where the separation of two Gaussian components is smaller. The degree of improvement can be quantitatively described by resolution, which is defined by:
\begin{equation}\label{eq:resolution}
	R=\dfrac{2 \times (t_2-t_1)}{1.70 \times (W_{1,h/2}+W_{2,h/2})}
\end{equation}
where $t_1$, $t_2$ are the peak positions of the two components, and $W_{1,h/2}$, $W_{2,h/2}$ are the full widths at half maximum (FWHM) of the two components, respectively. We get these parameters by fitting these distributions by two Gaussian components. The resolutions of SGRBs and LGRBs are $R_{T} \approx 0.49	$ based on $\log T_{90}$ and $R_{L} \approx 0.58$ based on $L$, respectively. The classification resolution has been improved with the new criterion.

\subsection{Another Possible Classification Scheme}

\begin{figure*}
	\begin{center}
		\includegraphics[width = 400 pt]{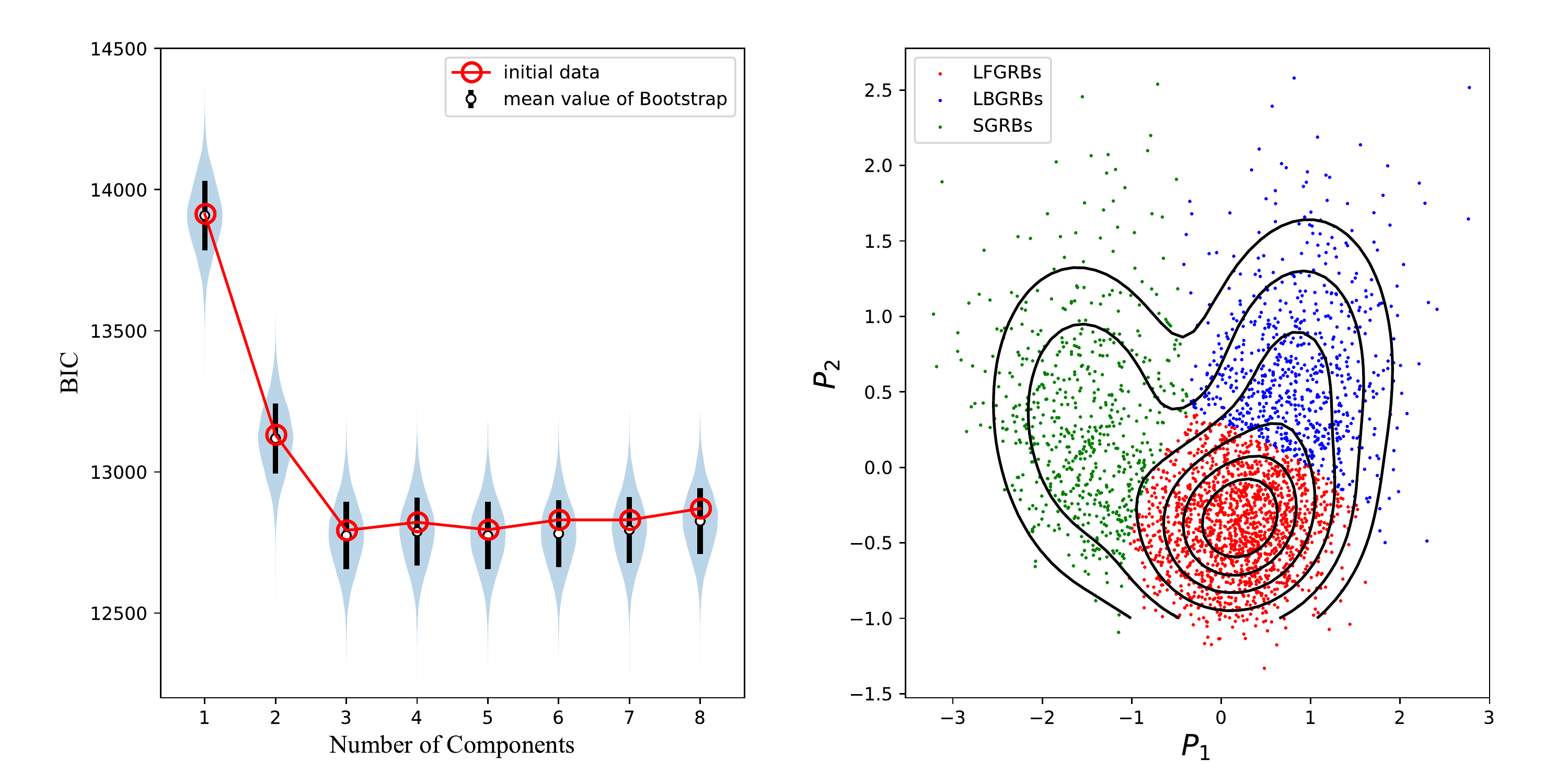}
		\caption{{\rm Left panel: } BIC score for different numbers of components. Red line with circles shows the BIC scores for different numbers of components derived from original data. The light-blue violins show probability distributions of BIC for each model, which derived from 10,000 times bootstrap sampling. The white dot with heavy black bar marks the mean values and 1$\sigma$ errors for different models. Obviously, the three-components model with the lowest BIC score is the best model in fitting the distribution of initial GRBs sample. {\rm Right panel: } classification of GRBs by three components of GMM. The contours represent equal-probability lines. Compared to Fig.\ref{fig3:LDA2class}, the green cluster roughly corresponds to SGRBs, red and blue clusters are roughly separated from LGRBs. }
		\label{fig4:BIC}
	\end{center}
\end{figure*}

\begin{figure}
	\begin{center}
		\includegraphics[width = 250 pt]{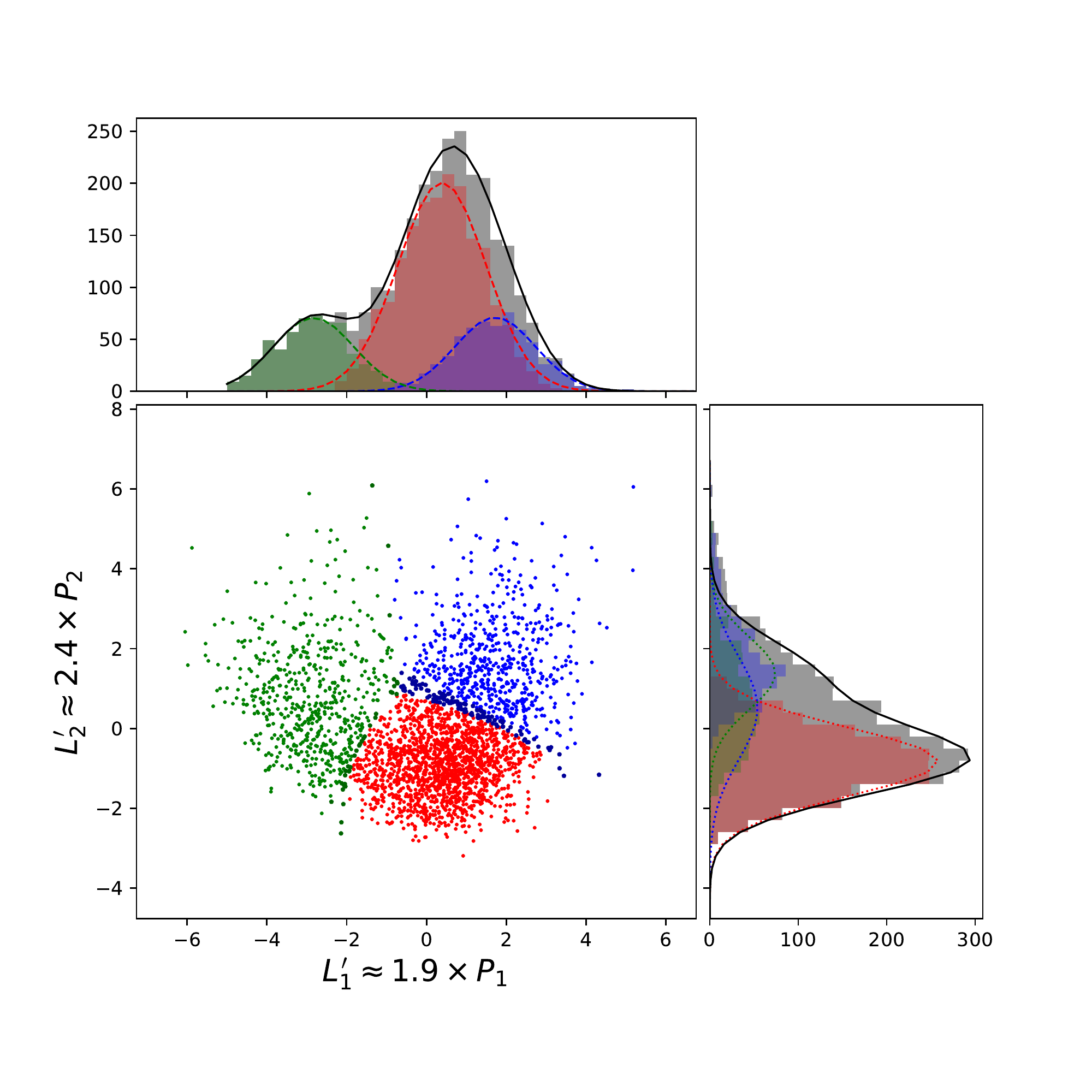}
		\caption{Reclassification and transferred to new coordinate by LDA. The axes ($L_{1}^{\prime}$ and $L_{2}^{\prime}$) are linearly proportional to the first and second principle components. The upper and right panels show the histogram of $L_{1}^{\prime}$ and $L_{2}^{\prime}$, respectively. The distribution of each subclass is fitted with a Gaussian function and the histogram of total sample is shown by a black line, which is the sum of the three Gaussian functions.}
		\label{fig5:3class}
	\end{center}
\end{figure}

Much research on GRB classification indicated to different models with different numbers of subclasses. The model selection can be performed with multicomponent GMMs using the Bayesian information criterion (BIC), which can effectively prevent the high model complexity caused by the high accuracy of the model. In order to improve the confidence of the results, we used the Bootstrap with replacement sampling method to extract 10,000 samples, and then performed the distribution of BIC scores. As shown in Fig.~\ref{fig4:BIC}, we performed from one- to eight-component models and found that a three-component model with the lowest BIC score is the optimal result. 

The three-component classification scheme is shown in the right panel of Fig.~\ref{fig4:BIC}. From the equal-probability lines, two protrusions extending from the highest density center cluster to two different directions that are caused by the other two clusters. LDA attempts to model the difference between the classes of data and it could project data to direction on which the different classes separated as much as possible while the same classes concentrated as much as possible. By this method the data could be transferred into new coordinate system, as shown in Fig.\ref{fig5:3class}, whose axes are obtained by LDA according Fisher criteria. The axes of new coordinate system are:
\begin{eqnarray}\label{eq:3classLDATransfor}
	(L_{1}^{\prime} \ L_{2}^{\prime}) &=& \begin{pmatrix} \nonumber
		P_1 \ P_2
	\end{pmatrix} \cdot  
	\begin{pmatrix}
		~1.877 & ~-0.005~\\
		~-0.010 & ~~2.399~\\
	\end{pmatrix} \\
	&\approx& \begin{pmatrix}
		1.9 P_1 \ \ 2.4 P_2
	\end{pmatrix}
\end{eqnarray}
Notice that the non-diagonal elements of the coordinate transfer matrix in Eq. (\ref{eq:3classLDATransfor}) are almost negligible; this means that the coordinate transfer hardly rotates the original coordinate system. It is an interesting coincidence that the first principal component is almost identical to the first direction of LDA, even though they are obtained through different criteria (maximum projection variance for PCA and Fisher criteria for LDA). This may be caused by physical conditions of different origin of GRBs.

Indeed, the highest density center gathers most of the long GRBs with lower luminosity (the red cluster) and one of the protrusions points to the short GRBs area (the green cluster); the other protrusions of blue cluster points to the long GRBs with higher luminosity. For convenience, we use short GRBs (SGRBs), long-bright GRBs (LBGRBs) and long-faint GRBs (LFGRBs) to refer to the three subclasses, although here short and long is no longer strictly identified by $T_{90}$.

The three subclasses are correspond to three clusters as shown in lower left panel of Fig.\ref{fig5:3class} whose axes are $L_{1}^{\prime}$ and $L_{2}^{\prime}$. The upper and right panels show the histogram of $L_{1}^{\prime}$ and $L_{2}^{\prime}$, respectively. The distribution of each subclass is fitted with a Gaussian function, the histogram of total sample is shown fitted by a black line, which is the sum of the three Gaussian function. One can easily see that the resolution from projection distribution on $L_{1}^{\prime}$ is better than the projection distribution on $L_{2}^{\prime}$.

\begin{figure*}
	\begin{center}
		\includegraphics[width = 450 pt]{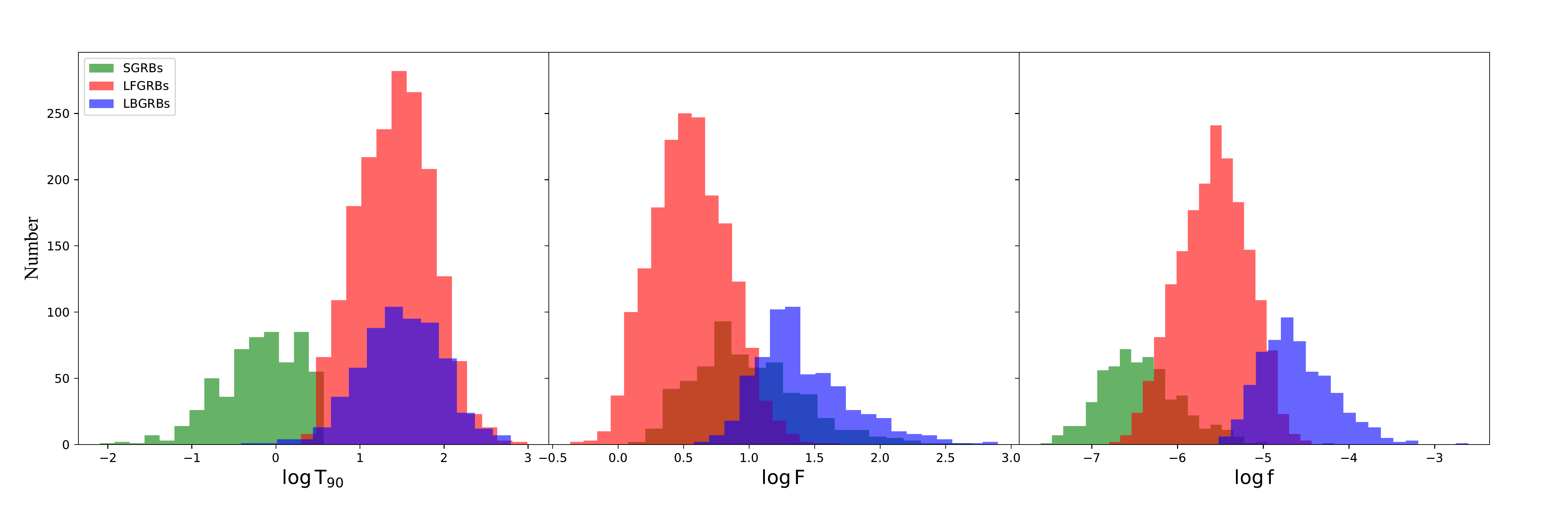}
		\caption{Distribution of $\log T_{90}, \log f $ and $\log F$ of the three subclasses. $T_{90}$ of SGRBs is statistically shorter than others, LBGRBs have statistically higher fluence and peak flux than LFGRBs, while their $T_{90}$ have similar distributions. }
		\label{fig6:characters}
	\end{center}
\end{figure*}

For further detail of the statistical characters of the three subclasses, we do the comparison of distribution of $\log T_{90}, \log f $ and $\log F$ of the three subclasses; the result is shown in Fig.\ref{fig6:characters}. Obviously, LBGRBs have greater statistical fluence and peak flux than LFGRBs, and $T_{90}$ of SGRBs is statistically shorter than others. 

\begin{figure}
	\begin{center}
		\includegraphics[width = 250 pt]{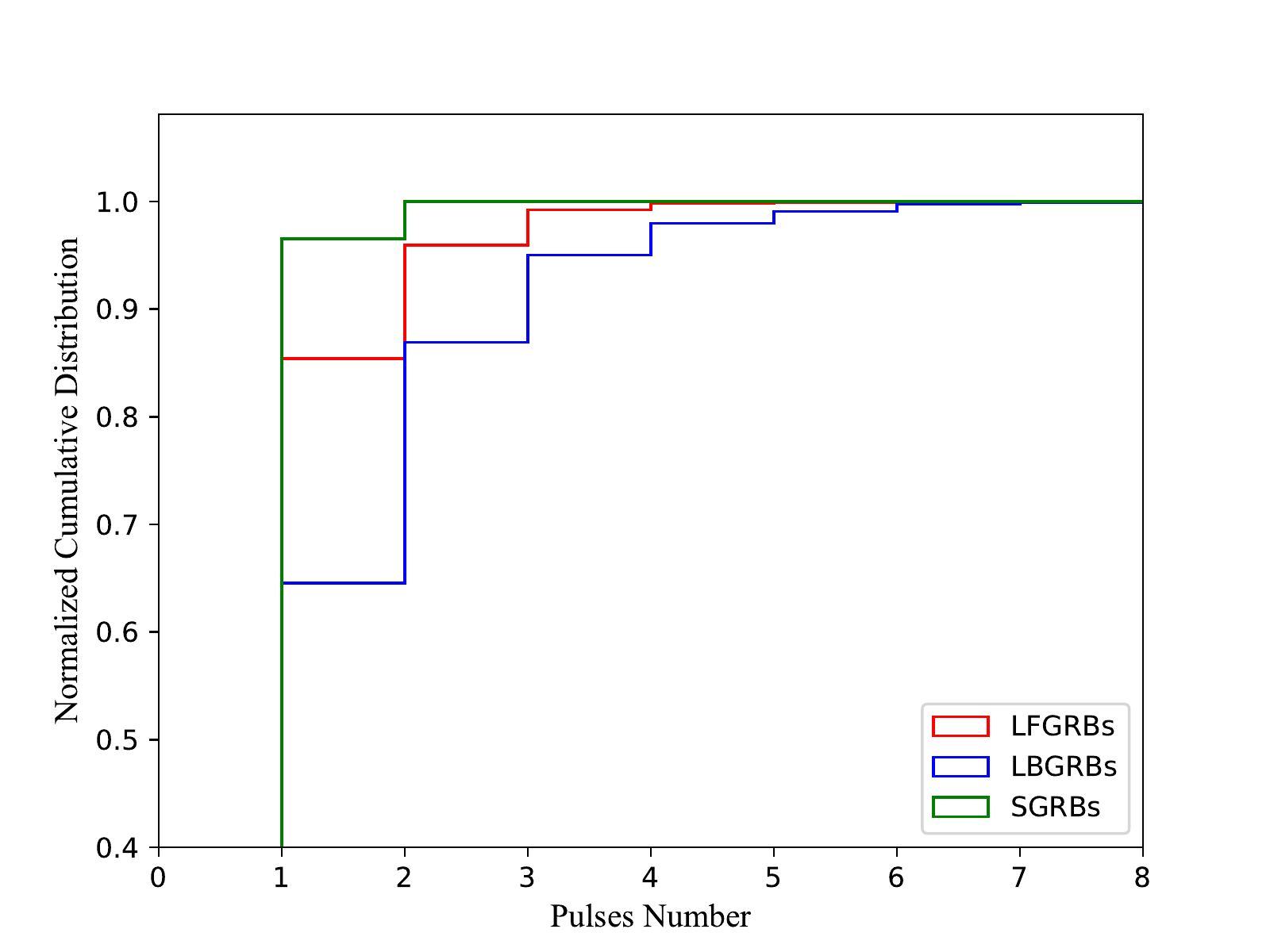}
		\caption{Normalized cumulative distribution of the number of pulses for three subclasses of GRBs. The KS-test shows the $P$ value between LBGRBs and LFGRBs is $6 \times 10^{-15}$, while $P$ values between LBGRBs and SGRBs is $2 \times 10^{-15}$, and the $P$ value between LFGRBs and SGRBs is $1\times 10^{-5}$.}
		\label{fig7:pulse}
	\end{center}
\end{figure}

A potentially relevant factor may be the number of GRB pulses. The normalized cumulative distribution of the number of pulses for three subclasses are shown in Fig.\ref{fig7:pulse}.  According to our classification results, more than $ 96.6\%$ of SGRBs have only one pulse, which for LFGRBs and LBGRBs are $85.1\%$ and $64.3\%$, respectively. The KS-test derived very small $P$ values, which indicates that three subclasses of GRBs have distinct distributions of the pulses numbers. LBGRBs statistically have more pulses than LGFRBs which may be the direct factor leading to the higher $f$ and $F$. However, it is not yet known whether the number of GRBs pulses involves different physical mechanisms. More in-depth theoretical research is still needed.

\section{ Simulation of the Correlation and Classification}
The connection between these parameters points to the similarity of each energy release process of GRBs. Under such rough assumptions, we can describe the process of GRB prompt emission by:
\begin{equation}\label{eq:locem}
	f^\prime=\bar e^\prime_{\rm ph} \cdot \bar F^\prime \cdot T^\prime _{\rm eff}
\end{equation}
where $ T^\prime _{\rm eff} $ represents the effective duration of each GRB,  $ \bar F^\prime $ and  $ f^\prime $ represent the average photon flux during the burst and the fluence (flux integrated over the burst duration), $ \bar e^\prime_{\rm ph} $ represents average energy of each photon, which can be considered equal for different bursts. All the prime marks indicate the parameters valued in local coordinate systems.
	
Considering the expansion effect of the universe, in the observer coordinate system, Eq.\ref{eq:locem} can be written as:
\begin{equation}\label{eq:obsem}
	\dfrac{f^\prime}{(1+z)}= \dfrac{\bar e^\prime_{\rm ph}}{(1+z)} \cdot \dfrac{\bar F^\prime}{(1+z)} \cdot T^\prime _{\rm eff}(1+z)
\end{equation}
Substitute the local variables to observer variables and, Eq.\ref{eq:obsem} can be written as:
\begin{equation}\label{eq:obseq}
	f = \dfrac{\bar e^\prime_{\rm ph}}{(1+z)} \cdot \bar F \cdot T_{\rm eff} 
\end{equation}
	
In view of the complexity of parameter definitions and measurements, we cannot directly measure $\bar F $ and $ T_{\rm eff} $. However, we can substitute them by common parameters according to $\bar F \propto F_{p}^{\alpha} $ and $ T_{eff} \propto T_{90}^{\beta}$ and then we have:
\begin{equation}\label{eq:precise_crl}
	f \propto (1+z)^{-1} \cdot F_{\rm p}^{\alpha} \cdot T_{90}^{\beta} 
\end{equation}
	
Since the measurement of red-shift is a strong bias on brightness, and most measured red-shift located in $1 \sim 3$, ignoring the redshift factor can greatly increase the statistical sample size.  So the upper correlation could be roughly expressed as:
\begin{equation}\label{eq:rough_crl}
	\log f \sim \alpha\log F_{\rm p}+ \beta\log T_{90} +\it const
\end{equation}
which is formally consistent with Eq.\ref{eq:3pfit}.

Based on the above discussion, we use the Monte Carlo (MC) method to simulate this correlation and classification by generating 3000 GRB light curves. The single-pulse profile of GRBs can be expressed by the function\citep{2005ApJ...627..324N,2009ApJ...705..372H,2012MNRAS.419.1650N}:
\begin{equation}\label{eq:lc_profile}
	I(t) = A \lambda \it e^{[-\tau_1/(t-t_{\rm s})-(t-t_s)/\tau_2]}
\end{equation}
where $A$ is the amplitude, $t$ is the time since trigger, $t_{\rm s}$ is the start time of the pulse, $\tau_1$ and $\tau_2$ are characteristics of the pulse rise
and decay, and $\lambda=exp[2(\tau_1/\tau_2)^{1/2}]$. Two parameters equivalent to $\tau_1$ and $\tau_2$ are $ \mu=\tau_1/\tau_2$ and $\tau_{\rm p}=\sqrt{\tau_1 \tau_2}$.

Eq.\ref{eq:lc_profile} gives the GRB prompt emission containing a single pulse. Based on this, we can get the multi-pulse GRB light curves:
\begin{equation}\label{eq:lc_multipulse_profile}
	I_n(t) =\sum_{i=1}^{n} A_i \lambda_i \it e^{[-\tau_{{\rm 1},i}/(t-t_{{\rm s},i})-(t-t_{{\rm s},i})/\tau_{{\rm 2},i}]}, ~~ n\ge 1
\end{equation}

The simulation procedure is summarized as follows: First, simulate a GRB prompt light curve by randomly inputting $A_i$, $\tau_{1,i}$, $\tau_{2,i}$, $t_{{\rm s},i}$ and $n$ into Eq.\ref{eq:lc_multipulse_profile}; second, calculate parameters ($T_{90}$, $f$, and $F$) from simulated light curve; finally, repeat the above steps by adjusting the initial parameter distributions, and make the result consistent with the real data. To this end, we need two sets of independently distributed parameters, one set is used to generate SGRBs, the other set is used to generate LGRBs which is divided into two categories according to the number of pulses. Finally, the following parameter distribution assumptions (not from real statistics) is adopted:
\begin{enumerate}
	\item $\tau_1$ and $\tau_2$: Since $\mu=\tau_1/\tau_2 $ and $\tau_{\rm p}=\sqrt{\tau_1 \tau_2}$ more intuitively reflect the characteristics of light curve profiles, we generate random $\mu $ and $\tau_{\rm p}$ instead of $\tau_1$ and $\tau_2$ directly. For LGRBs, $\mu \sim uniform(0.1,0.5) $ and $\log \tau_{\rm p} \sim N(0.7,0.5)$. For SGRBs, $\mu \sim uniform(0.1,0.5) $ and $\log \tau_{\rm p} \sim N(-1,0.5)$.
	\item $A$: Intuitively, peak of a single pulse. For SGRBs, $\log A \sim N(0.6,0.5) $ and for SGRBs $\log A \sim N(1,0.5)$.
	\item $t_{\rm s}$: For SGRBs, $t_{\rm s} \sim N(2,12) $ and for SGRBs $ t_{\rm s} \sim N(0.5,1.5)$.
	\item $n$: Number of pulses for GRBs modeled by integer of Erlang distribution $ Erlang(1,1)$.
\end{enumerate}

In order to make the GBM data consistent with the simulated data, two other parameters should be adopted according to Eq.\ref{eq:obseq}, i.e. average energy of each photon($\bar e^\prime_{ph}$) and red-shift ($z$). The average energy of each photon for LGRBs and SGRBs are $1.2 \times 10^{-8}~\rm erg~photon^{-1}$ and $3 \times 10^{-8}~\rm erg~photon^{-1}$. Red-shift generated for each GRBs randomly according to the probability density function $ Erlang(2,1)$\citep{2011ApJ...738...19S}.

\begin{figure*}[t!]
	\centering
	\begin{minipage}[t]{0.4\textwidth}
		\centering
		\includegraphics[width=\textwidth]{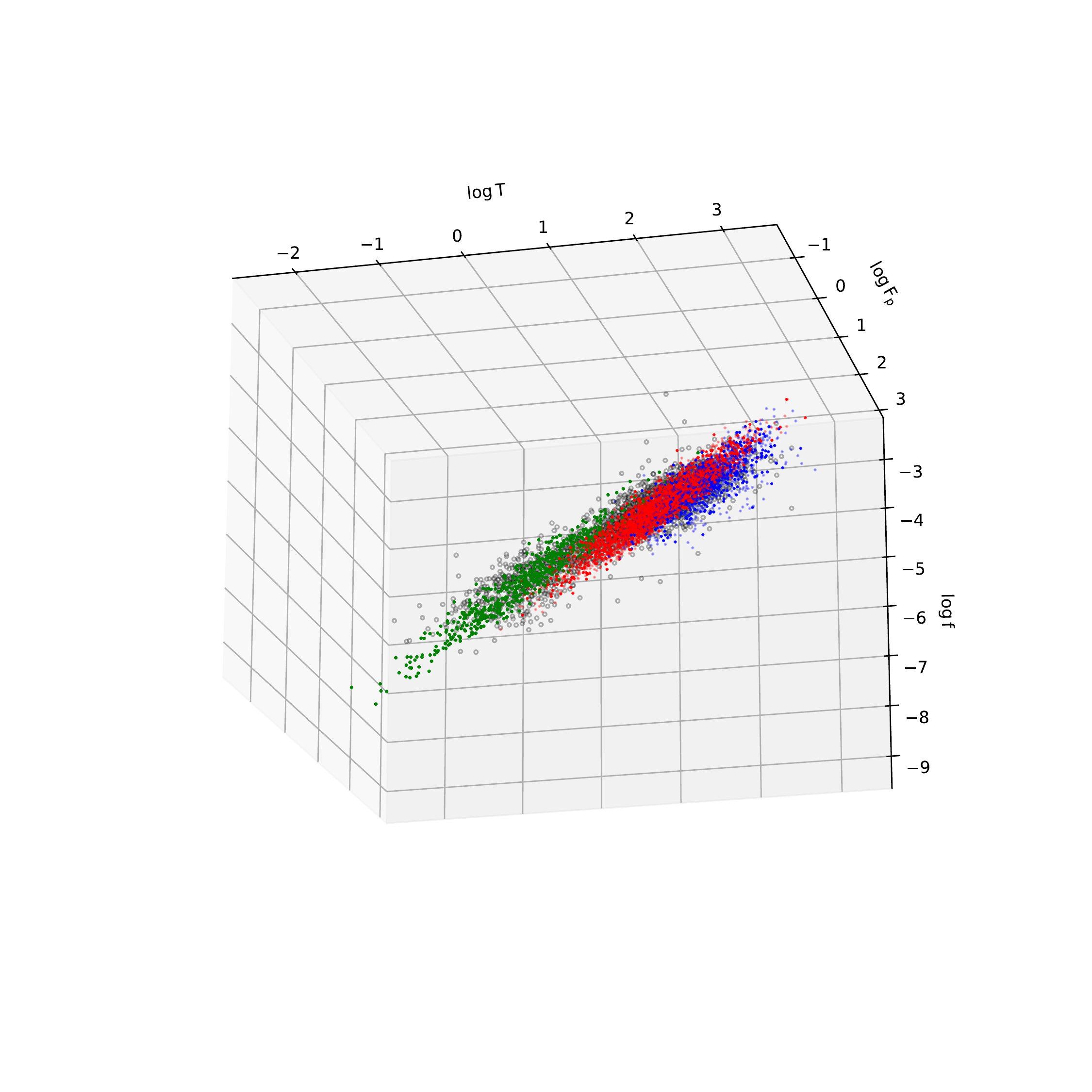}
	\end{minipage}
	\begin{minipage}[t]{0.4\textwidth}
		\centering
		\includegraphics[width=\textwidth]{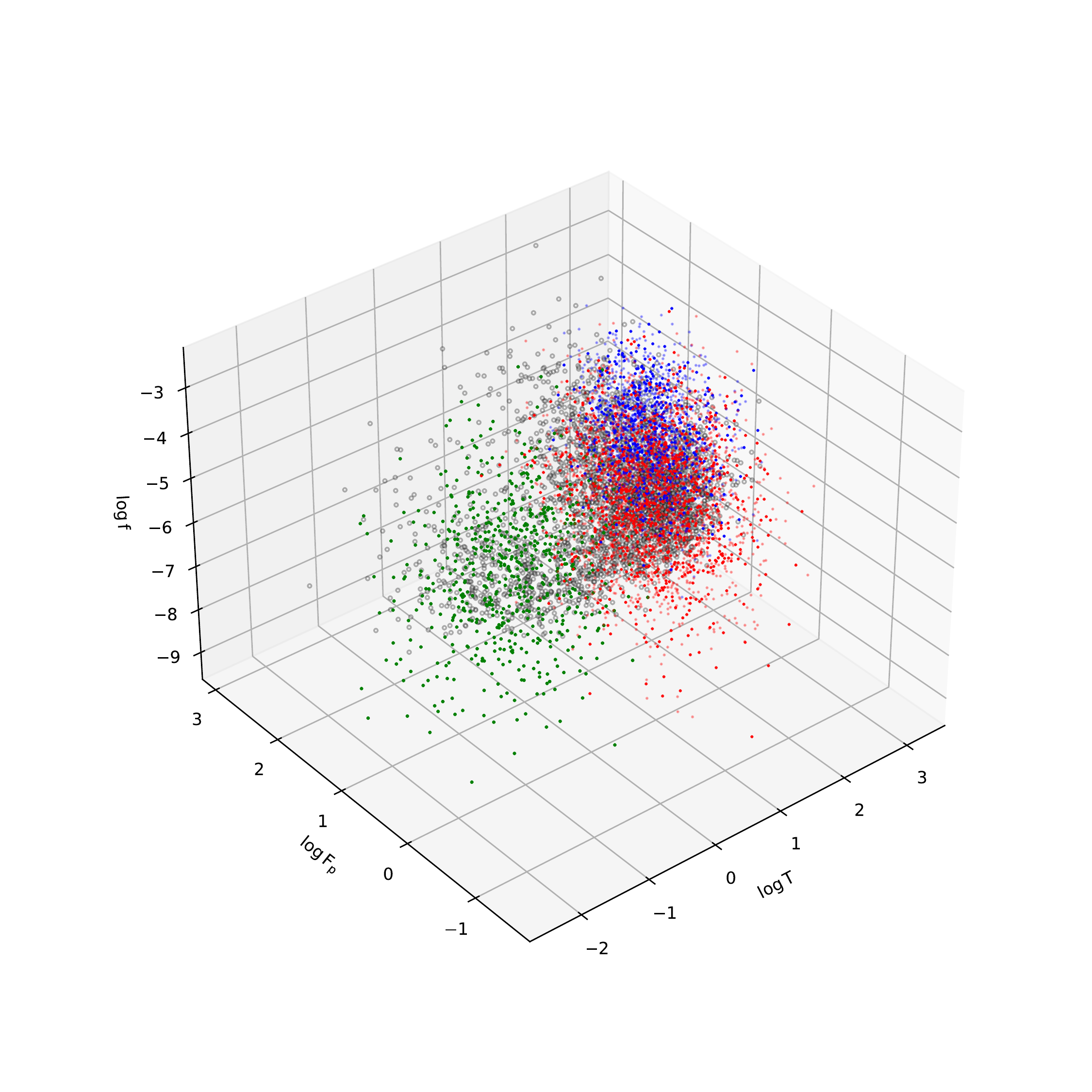}
	\end{minipage}
	\caption{Comparison of the distribution of GBM data and simulated data from different angle of view. Black points represent real GRBs from GBM data, green points represent simulated SGRBs and red (blue) points represent simulated long GRBs with single (multiple) pulse(s).}
	\label{fig8:MC_simu}
\end{figure*}

In the left panel of Fig.\ref{fig8:MC_simu}, as comparison, the black circles represent real GRB data from GBM, green dots represent simulated SGRBs and red (blue) dots represent simulated long GRBs with single (multiple) pulse(s), simulated data perfectly satisfies the same correlation as real data. From the perspective perpendicular to the left panel, the right panel of Fig.\ref{fig8:MC_simu} shows the most discrete distribution of all GRBs. In addition to SGRB as an independent cluster, the cluster of LGRBs with multiple pulses also deviates from the LGRBs with a single pulse. This result confirms the GMM classification scheme in Fig.\ref{fig4:BIC} and the pulse statistics of each subclasses reflected in Fig.\ref{fig7:pulse}. 

There are a small number of simulated samples that exceed the data distribution at the location of low peak fluxes and fluences, which are most likely to correspond to GRBs that were not observed by the bias effect.

In summary, the difference between LFGRBs and LBGRBs is ostensibly derived from the light curve model. Statistically, those GRBs with multiple pulses have higher peak fluxes and fluences. As for the essential difference between the two subclasses of LGRBs, more dimensional data is needed for research.

\section{Summary and discussion}\label{sec:Summary}

In this work, we carry out some analysis on various distributions of GRBs by the data mining method. We first extract a tight correlation by using PCA on three-dimensional data. The third principle component contributes the least and has ignorable variance ratio, which indicates the intrinsic consistency of GRBs. 

Based on the first and second principle ($P_1$ and $P_2$) components, which contain most of the variance, we composed classification analysis of two- and three-subclasses schemes. 
 
In order to better distinguish SGRBs and LGRBs, we reduced the dimensionality of the data by LDA and found a compound parameter ($L$) that provided a higher resolution between SGRBs and LGRBs. The resolution of the two subclasses measured by $L$ is 0.58, it is obviously improved from 0.49, which is given by $T_{90}$. For the three-components model, we found the first principle component $P_1$ is almost identical to the first direction $L_{1}^{\prime}$ of LDA, which is obtained through Fisher criteria with the purpose of maximizing the covariance between classes and the minimum covariance within classes. This may be purely a coincidence in the data distribution or may be caused by different physical conditions where GRBs originated.

In general, the only uncontroversial and conclusive subclass of GRBs is SGRBs. Although the distinguish at the parameter boundary is controversial, there is a lot of evidence for its existence, which is widely recognized. Regarding LGRBs and their subclasses, we still need more information introduced by other independent parameters to help solve this problem.

Regardless of the number of parameters, PCA and LDA can be used to reduce the dimensionality to one or two if the number of subclasses is not more than three. It is easier to visualize data distribution in low-dimensional space.

The three subclasses classification scheme has been discussed in different aspects\citep{2009MNRAS.392...91V,2017arXiv170805668B,2019JPhCS1400b2010S,2019Ap&SS.364..105H}. Classifying LGRBs based on hardness ratio, peak flux, or luminosity have similar conclusions as our analysis. However, such a classification scheme is still not sufficiently supported by the combination of theory and data.

So far, the existence and origin of SGRBs is widely recognized in that the classification based on $T_{90}$ points to the difference in the physical scales of the GRB progenitors. The influence of other physical parameters, such as Lorentz factor ($\Gamma$), the fraction of energy of shocked electrons ($\epsilon_{\rm e}$), or magnetic field ($\epsilon_{\rm B}$) on radiation is more complicated. Complex physical information is coupled in many observed parameters; this indicates that it may be possible to derive the physical information of the source through the combination of observed parameters. It may be helpful to introduce more independent parameters containing more physical information, and the analysis method of this work is easy to extend by introducing new independent variables. However, many parameters of SGRBs or LFGRBs are difficult to measure, such as the red-shift and related quantities whose measurements require higher brightness. 

As a supplement to pure data analysis, we generate a sample by the MC method based on an analytical light curve model \citep{2005ApJ...627..324N,2009ApJ...705..372H,2012MNRAS.419.1650N}, and then reproduce the strong correlation and clusters distribution. The simulation confirms that, under the unified light curve model, there must be a strong correlation as in Eq \ref{eq:3pfit}. And finally we found that the light curves composed by multiple pulses lead to statistically higher fluence and peak flux. This also explains why the LGRBs are composed of two subclusters, and consistent with the pulse distribution shown in Fig.\ref{fig7:pulse}.

Another point worth noting is that, considering the definition of statistical parameters, if the research object is replaced with single GRB pulses, LBGRBs and LFGRBs may be unified into one category. 
And there are still other possibilities, such as the existence of fake LFGRBs that appeared purely due to background fluctuations. The fake triggers caused by background fluctuations are concentrated in low-brightness areas, which increased the proportion of single-pulse GRBs. If this part is removed, LFGRBs may have similar distributions of number of pulses as LBGRBs, ie., in that case, the necessity to divide LGRBs into two subclasses still needs more in-depth discussion.

\section*{Acknowledgments}

This work was funded by the National Natural Science Foundation of China (No. 11803006, 11903012); Science and Technology Project of Hebei Education Department (No. QN2018036) and Hebei NSF (No. A2019205166).


\begin{thebibliography}{}
	\bibitem[Berger et al.(2005)]{2005Natur.438..988B} Berger, E., Price, P.~A., Cenko, S.~B., et al.\ 2005, \dataset[\color{blue} \nat, 438, 988. ]{\doi{10.1038/nature04238}}
	\bibitem[Bhave et al.(2017)]{2017arXiv170805668B} Bhave, A., Kulkarni, S., Desai, S., et al.\ 2017, \dataset[\color{blue}arXiv:1708.05668]{https://arxiv.org/abs/1708.05668}
	\bibitem[Bloom et al.(2002)]{2002AJ....123.1111B} Bloom, J.~S., Kulkarni, S.~R., \& Djorgovski, S.~G.\ 2002, \dataset[\color{blue} \aj, 123, 1111. ]{\doi{10.1086/338893}}
	\bibitem[Bloom et al.(2006)]{2006ApJ...638..354B} Bloom, J.~S., Prochaska, J.~X., Pooley, D., et al.\ 2006, \dataset[\color{blue} \apj, 638, 354. ]{\doi{10.1086/498107}}
	\bibitem[Chattopadhyay \& Maitra(2017)]{2017MNRAS.469.3374C} Chattopadhyay, S. \& Maitra, R.\ 2017, \dataset[\color{blue} \mnras, 469, 3374. ]{\doi{10.1093/mnras/stx1024}}
	\bibitem[Chattopadhyay \& Maitra(2018)]{2018MNRAS.481.3196C} Chattopadhyay, S. \& Maitra, R.\ 2018, \dataset[\color{blue} \mnras, 481, 3196. ]{\doi{10.1093/mnras/sty1940}}
	\bibitem[Dinov, I.D. (2008)]{Dinov2008} Dinov, I. D. \ 2008  \dataset[\color{blue} UCLA: Statistics Online Computational Resource.]{https://escholarship.org/uc/item/1rb70972}
	\bibitem[Eichler et al.(1989)]{1989Natur.340..126E} Eichler, D., Livio, M., Piran, T., et al.\ 1989, \dataset[\color{blue} \nat, 340, 126. ]{\doi{10.1038/340126a0}}
	\bibitem[Fisher, R.A. (1936)]{Fisher1936} Fisher, R.A. \ 1936 Annals of Eugenics, 7, 179
	\bibitem[Fong et al.(2011)]{2011ApJ...730...26F} Fong, W., Berger, E., Chornock, R., et al.\ 2011, \dataset[\color{blue} \apj, 730, 26. ]{\doi{10.1088/0004-637X/730/1/26}}
	\bibitem[Fong \& Berger(2013)]{2013ApJ...776...18F} Fong, W. \& Berger, E.\ 2013, \dataset[\color{blue} \apj, 776, 18. ]{\doi{10.1088/0004-637X/776/1/18}}
	\bibitem[Fox et al.(2005)]{2005Natur.437..845F} Fox, D.~B., Frail, D.~A., Price, P.~A., et al.\ 2005, \dataset[\color{blue} \nat, 437, 845. ]{\doi{10.1038/nature04189}}
	\bibitem[Fruchter et al.(2006)]{2006Natur.441..463F} Fruchter, A.~S., Levan, A.~J., Strolger, L., et al.\ 2006, \dataset[\color{blue} \nat, 441, 463. ]{\doi{10.1038/nature04787}}
	\bibitem[Guoshen Yu et al.(2012)]{2012ITIP...21.2481G} Guoshen Yu, Sapiro, G., \& Mallat, S.\ 2012, \dataset[\color{blue} IEEE Transactions on Image Processing, 21, 2481. ]{\doi{10.1109/TIP.2011.2176743}}
	\bibitem[Hakkila et al.(2000)]{2000ApJ...538..165H} Hakkila, J., Haglin, D.~J., Pendleton, G.~N., et al.\ 2000, \dataset[\color{blue} \apj, 538, 165. ]{\doi{10.1086/309107}}
	\bibitem[Hakkila \& Nemiroff(2009)]{2009ApJ...705..372H} Hakkila, J. \& Nemiroff, R.~J.\ 2009, \dataset[\color{blue} \apj, 705, 372. ]{\doi{10.1088/0004-637X/705/1/372}}
	\bibitem[Horv{\'a}th(1998)]{1998ApJ...508..757H} Horv{\'a}th, I.\ 1998, \dataset[\color{blue} \apj, 508, 757. ]{\doi{10.1086/306416}}
	\bibitem[Horv{\'a}th et al.(2010)]{2010ApJ...713..552H} Horv{\'a}th, I., Bagoly, Z., Bal{\'a}zs, L.~G., et al.\ 2010, \dataset[\color{blue} \apj, 713, 552. ]{\doi{10.1088/0004-637X/713/1/552}}
	\bibitem[Horv{\'a}th et al.(2019)]{2019Ap&SS.364..105H} Horv{\'a}th, I., Hakkila, J., Bagoly, Z., et al.\ 2019, \dataset[\color{blue} \apss, 364, 105. ]{\doi{10.1007/s10509-019-3585-1}}
	\bibitem[Hotelling, H. (1933a)]{Hotelling1933a} Hotelling, H. \ 1933a. J. Educ. Psychol., 24, 417
	\bibitem[Hotelling, H. (1933b)]{Hotelling1933b} Hotelling, H. \ 1933b. J. Educ. Psychol., 24, 498
	\bibitem[Klebesadel et al.(1973)]{1973ApJ...182L..85K} Klebesadel, R.~W., Strong, I.~B., \& Olson, R.~A.\ 1973, \dataset[\color{blue} \apjl, 182, L85. ]{\doi{10.1086/181225}}
	\bibitem[Kobulnicky \& Kewley(2004)]{2004ApJ...617..240K} Kobulnicky, H.~A. \& Kewley, L.~J.\ 2004, \dataset[\color{blue} \apj, 617, 240. ]{\doi{10.1086/425299}}
	\bibitem[Kouveliotou et al.(1993)]{1993ApJ...413L.101K} Kouveliotou, C., Meegan, C.~A., Fishman, G.~J., et al.\ 1993, \dataset[\color{blue} \apjl, 413, L101. ]{\doi{10.1086/186969}}
	\bibitem[Kulkarni et al.(2000)]{2000AIPC..526..277K} Kulkarni, S.~R., Berger, E., Bloom, J.~S., et al.\ 2000, \dataset[\color{blue} Gamma-ray Bursts, 5th Huntsville Symposium, 526, 277. ]{\doi{10.1063/1.1361549}}
	\bibitem[Leibler \& Berger(2010)]{2010ApJ...725.1202L} Leibler, C.~N. \& Berger, E.\ 2010, \dataset[\color{blue} \apj, 725, 1202. ]{\doi{10.1088/0004-637X/725/1/1202}}
	\bibitem[Li et al.(2011)]{2011MNRAS.412.1473L} Li, W., Chornock, R., Leaman, J., et al.\ 2011, \dataset[\color{blue} \mnras, 412, 1473. ]{\doi{10.1111/j.1365-2966.2011.18162.x}}
	\bibitem[Li et al.(2016)]{2016ApJS..227....7L} Li, Y., Zhang, B., \& L{\"u}, H.-J.\ 2016, \dataset[\color{blue} \apjs, 227, 7. ]{\doi{10.3847/0067-0049/227/1/7}}
	\bibitem[L{\"u} et al.(2010)]{2010ApJ...725.1965L} L{\"u}, H.-J., Liang, E.-W., Zhang, B.-B., et al.\ 2010, \dataset[\color{blue} \apj, 725, 1965. ]{\doi{10.1088/0004-637X/725/2/1965}}
	\bibitem[MacFadyen \& Woosley(1999)]{1999ApJ...524..262M} MacFadyen, A.~I. \& Woosley, S.~E.\ 1999, \dataset[\color{blue} \apj, 524, 262. ]{\doi{10.1086/307790}}
	\bibitem[Marchetti et al.(2017)]{2017A&A...600A..54M} Marchetti, A., Garilli, B., Granett, B.~R., et al.\ 2017, \dataset[\color{blue} \aap, 600, A54. ]{\doi{10.1051/0004-6361/201630249}}
	\bibitem[Mukherjee et al.(1998)]{1998ApJ...508..314M} Mukherjee, S., Feigelson, E.~D., Jogesh Babu, G., et al.\ 1998, \dataset[\color{blue} \apj, 508, 314. ]{\doi{10.1086/306386}}
	\bibitem[Narayan et al.(1992)]{1992ApJ...395L..83N} Narayan, R., Paczynski, B., \& Piran, T.\ 1992, \dataset[\color{blue} \apjl, 395, L83. ]{\doi{10.1086/186493}}
	\bibitem[Nemiroff(2012)]{2012MNRAS.419.1650N} Nemiroff, R.~J.\ 2012, \dataset[\color{blue} \mnras, 419, 1650. ]{\doi{10.1111/j.1365-2966.2011.19838.x}}
	\bibitem[Norris et al.(2005)]{2005ApJ...627..324N} Norris, J.~P., Bonnell, J.~T., Kazanas, D., et al.\ 2005, \dataset[\color{blue} \apj, 627, 324. ]{\doi{10.1086/430294}}
	\bibitem[Paczynski(1991)]{1991AcA....41..257P} Paczynski, B.\ 1991, \actaa, 41, 257
	\bibitem[Pearson (1901)]{Pearson1901} Pearson, K. \ 1901. \dataset[\color{blue} Phil. Mag. Ser.6, 2, 559. ]{\doi{10.1080/14786440109462720}}
	\bibitem[Savaglio et al.(2009)]{2009ApJ...691..182S} Savaglio, S., Glazebrook, K., \& Le Borgne, D.\ 2009, \dataset[\color{blue} \apj, 691, 182. ]{\doi{10.1088/0004-637X/691/1/182}}
	\bibitem[Shao et al.(2011)]{2011ApJ...738...19S} Shao, L., Dai, Z.-G., Fan, Y.-Z., et al.\ 2011, \dataset[\color{blue} \apj, 738, 19. ]{\doi{10.1088/0004-637X/738/1/19}}
	\bibitem[Scargle et al.(2013)]{2013ApJ...764..167S} Scargle, J.~D., Norris, J.~P., Jackson, B., et al.\ 2013, \dataset[\color{blue} \apj, 764, 167. ]{\doi{10.1088/0004-637X/764/2/167}}
	\bibitem[{\v{S}}koda et al.(2016)]{2016arXiv161207549S} {\v{S}}koda, P., Shakurova, K., Koza, J., et al.\ 2016, \dataset[\color{blue} arXiv:1612.07549]{https://arxiv.org/abs/1612.07549}
	\bibitem[Svinkin et al.(2019)]{2019JPhCS1400b2010S} Svinkin, D.~S., Aptekar, R.~L., Golenetskii, S.~V., et al.\ 2019, \dataset[\color{blue} Journal of Physics Conference Series, 1400, 022010. ]{\doi{10.1088/1742-6596/1400/2/022010}}
	\bibitem[Tarnopolski(2019a)]{2019ApJ...870..105T} Tarnopolski, M.\ 2019a, \dataset[\color{blue} \apj, 870, 105. ]{\doi{10.3847/1538-4357/aaf1c5}}
	\bibitem[Tarnopolski(2019b)]{2019ApJ...887...97T} Tarnopolski, M.\ 2019b, \dataset[\color{blue} \apj, 887, 97. ]{\doi{10.3847/1538-4357/ab4fe6}}
	\bibitem[Tremonti et al.(2004)]{2004ApJ...613..898T} Tremonti, C.~A., Heckman, T.~M., Kauffmann, G., et al.\ 2004, \dataset[\color{blue} \apj, 613, 898. ]{\doi{10.1086/423264}}
	\bibitem[Ukwatta et al.(2016)]{2016MNRAS.458.3821U} Ukwatta, T.~N., Wo{\'z}niak, P.~R., \& Gehrels, N.\ 2016, \dataset[\color{blue} \mnras, 458, 3821. ]{\doi{10.1093/mnras/stw559}}
	\bibitem[Virgili et al.(2009)]{2009MNRAS.392...91V} Virgili, F.~J., Liang, E.-W., \& Zhang, B.\ 2009, \dataset[\color{blue} \mnras, 392, 91. ]{\doi{10.1111/j.1365-2966.2008.14063.x}}
	\bibitem[von Kienlin et al.(2020)]{2020ApJ...893...46V} von Kienlin, A., Meegan, C. A., Paciesas, W. S., et al.\ 2020, \dataset[\color{blue} \apj, 893, 46 ]{\doi{10.3847/1538-4357/ab7a18}}
	\bibitem[Woosley \& Bloom(2006)]{2006ARA&A..44..507W} Woosley, S.~E. \& Bloom, J.~S.\ 2006, \dataset[\color{blue} \araa, 44, 507. ]{\doi{10.1146/annurev.astro.43.072103.150558}}
	\bibitem[Zhang et al.(2017)]{2017ApJ...844...55Z} Zhang, S., Jin, Z.-P., Zhang, F.-W., et al.\ 2017, \dataset[\color{blue} \apj, 844, 55. ]{\doi{10.3847/1538-4357/aa7aa7}}

\end{thebibliography}
\end{document}